\documentclass[a4paper,10pt,twoside]{cpc-hepnp}

\usepackage{multicol}
\usepackage{siunitx}
\usepackage{graphicx}
\usepackage{booktabs}
\usepackage{amssymb,bm,mathrsfs,bbm,amscd}
\usepackage[tbtags]{amsmath}
\usepackage{lastpage}
\usepackage[utf8]{inputenc}
\usepackage{CJK}
\usepackage[colorlinks=true,linkcolor=blue,urlcolor=blue,citecolor=blue]{hyperref}

\begin{document}
\begin{CJK*}{UTF8}{gbsn}

\title{Performance Evaluation of BPM System in SSRF Using PCA Method\thanks{Work supported by National Natural Science Foundation of China (11075198)}}

\author{%
      CHEN Zhi-Chu(陈之初)$^{1,2;1)}$\email{chenzhichu@sinap.ac.cn}%
\quad LENG Yong-Bin(冷用斌)$^{1,2;2)}$\email{lengyongbin@sinap.ac.cn, corresponding author.}%
\\
\quad YAN Ying-Bing(阎映炳)$^{1,2}$
\quad YUAN Ren-Xian(袁任贤)$^{1,2}$
\quad LAI Long-Wei(赖龙伟)$^{1,2}$
}
\maketitle

\end{CJK*}

\address{%
$^1$ Shanghai Institute of Applied Physics, Chinese Academy of Sciences, Shanghai 201800, China\\
$^2$ Shanghai Synchrotron Radiation Facility, Chinese Academy of Sciences, Shanghai 201203, China\\
}

\begin{abstract}
The beam position monitor~(BPM) system is of most importance in a light source. The capability of the BPM depends on the resolution of the system. The traditional standard deviation on the raw data method merely gives the upper limit of the resolution. Principal component analysis~(PCA) had been introduced in the accelerator physics and it could be used to get rid of the actual signals. Beam related informations were extracted before the evaluation of the BPM performance. A series of studies had been made in Shanghai Synchrotron Radiation Facility~(SSRF) and PCA was proved as an effective and robust method in the performance evaluations of our BPM system.
\end{abstract}




\begin{multicols}{2}

\section{Introduction}

As the key beam diagnostics tool, BPM systems are widely equipped in all kinds of accelerators and are being used in daily operation and machine study. The BPM system couldn't detect the beam dynamics which have smaller amplitudes than its resolution. The resolution of the BPM system should therefore determine the boundary of all measurable beam dynamics during machine study or daily operation. Demands for the precise measurement of the resolution of the BPM system arose when fine beam dynamics studies were proposed.

The performance of BPM electronics itself can be easy evaluated in the lab test by using the virtual beam signals simulated by an RF signal source and the power splitters. In this case the standard deviation of the raw position readings reflects the position resolution directly. However for a practical BPM system, which includes not only electronics but also signal cables, pickup electrodes and mechanical supports, working with real beams, it is a hard task to evaluate the position resolution precisely. The standard deviation of the raw position readings is hardly useful since the support vibration and the electronics noise are mixed in the real beam movement. A more powerful analyzing toolkits is required.

PCA, which is introduced in the particle accelerator field as the model-independent analysis~(MIA), is an excellent candidate for this job by separating real beam motion from raw data matrix via correlation analyzing.

\section{BPM System in the SSRF Ring}

The locations of the BPMs have been carefully selected. Each one of the total 20 cells was allocated 7 BPMs: 2 insertion-device~(ID) BPMs at the head and the tail of each cell and 5 arc BPMs at other positions. The ID BPMs were installed at the entrance or exit of a drift space while the arc BPMs were separated in the bending areas.

The commercial electronics used in the BPM system is called the ``Libera Electron.'' A more advanced version called the ``Libera Brilliance'' was released several years after the commissioning of SSRF. The upgrading has taken 3 steps since then. A detailed list of the procedure is shown in Table~\ref{tab:upgrade-details}. The 140 BPMs in the storage ring are designed for slightly different purposes, so the upgradings were rather complex.

\begin{center}\begin{minipage}{\hsize}
\tabcaption{\label{tab:upgrade-details}The course of the BPM electronics upgrading.}
\footnotesize
\begin{tabular*}{\hsize}{l@{\extracolsep{\fill}}ll}
\toprule
Milestone & Libera Brilliance Locations & Number \\
\hline
spring, 2009 & none & 0 \\[8pt]
summer, 2010 & $\left[ \frac54n \right] \times7+3$ & 15 \\[8pt]
summer, 2011 & $\left[ \frac59(n+1) \right]\times3+\left[ \frac59n \right]\times4+3$ & 35 \\[8pt]
summer, 2012 & $\left[ 6-\frac1{(n-1.42)^2}+n \right]\times7-2$ & 15 \\
             & and $\left[ \frac{(n+1)}2 \right]+\left[ \frac n2 \right]\times6$ & 40 \\
\bottomrule
\end{tabular*}
\end{minipage}\end{center}

The formulae in Table~\ref{tab:upgrade-details}, which indicate the IDs of the BPMs whose electronics was the Libera Brilliance, are merely mathematical expressions. They are used here to reduce the input and to avoid the waste of space, rather than to imply the motives of the upgradings or other meanings. Fortunately, these upgradings did not stick to a single BPM category and less bias would be involved in the results.

\section{Principle Component Analysis}

\subsection{Basic concept}

The BPM data matrix can be decomposed into 3 terms by using SVD:
\begin{equation}\label{eq:SVD}
B = U S V^\dagger ,
\end{equation}
where $U$~(also called the temporal matrix) and $V$~(called the spatial matrix) are unitary, and $S$~(called the singular value matrix) is diagonal. Equation \eqref{eq:SVD} could also be written as the following form by using the Einstein summation convention:
\begin{equation}\label{eq:SVD-Einstein}
b_n^t = u_k^t s_k^k v_n^k ,
\end{equation}
where $u_k$ is the $k$th column vector of $U$, $v^k$ the $k$th row vector of $V$ and $s_k^k$ the $k$th diagonal element of $S$. In this way, it is clear that the final signal is the superposition of all the linear-independent bases, temporally and spatially. Based on the theory of PCA, the bases of the temporal matrix, $u_k$'s, correspond to all the physical modes of the BPM system, the bases of the spatial matrix, $v^k$'s, the corresponding normalized distribution functions of the BPMs along the ring, and the singular values, $s_k^k$'s, the corresponding deviations of the physical modes.\cite{Irwin-Wang:1999:PRL,Wang:2004:PRSTAB,Wang-Calaga:2004:EPAC,Wang:2004:EPAC,Calaga-Tomas:2004:PRSTAB}

\subsection{Noise evaluation method}

The diagonal elements of the singular matrix $S$ provides an estimate of the modes. As a typical result of the SVD, figure~\ref{fig:typical-singular-values} shows that there is a floor of these singular values and about a dozen modes are separable from the floor. These modes could be useful to study the beam dynamics and the floor might reflect the average noise of the BPM system.\cite{Wang-Borland:2001:PAC,Wolski-Ross:2002:EPAC}

\begin{center}\begin{minipage}{\hsize}
\includegraphics[width=\hsize]{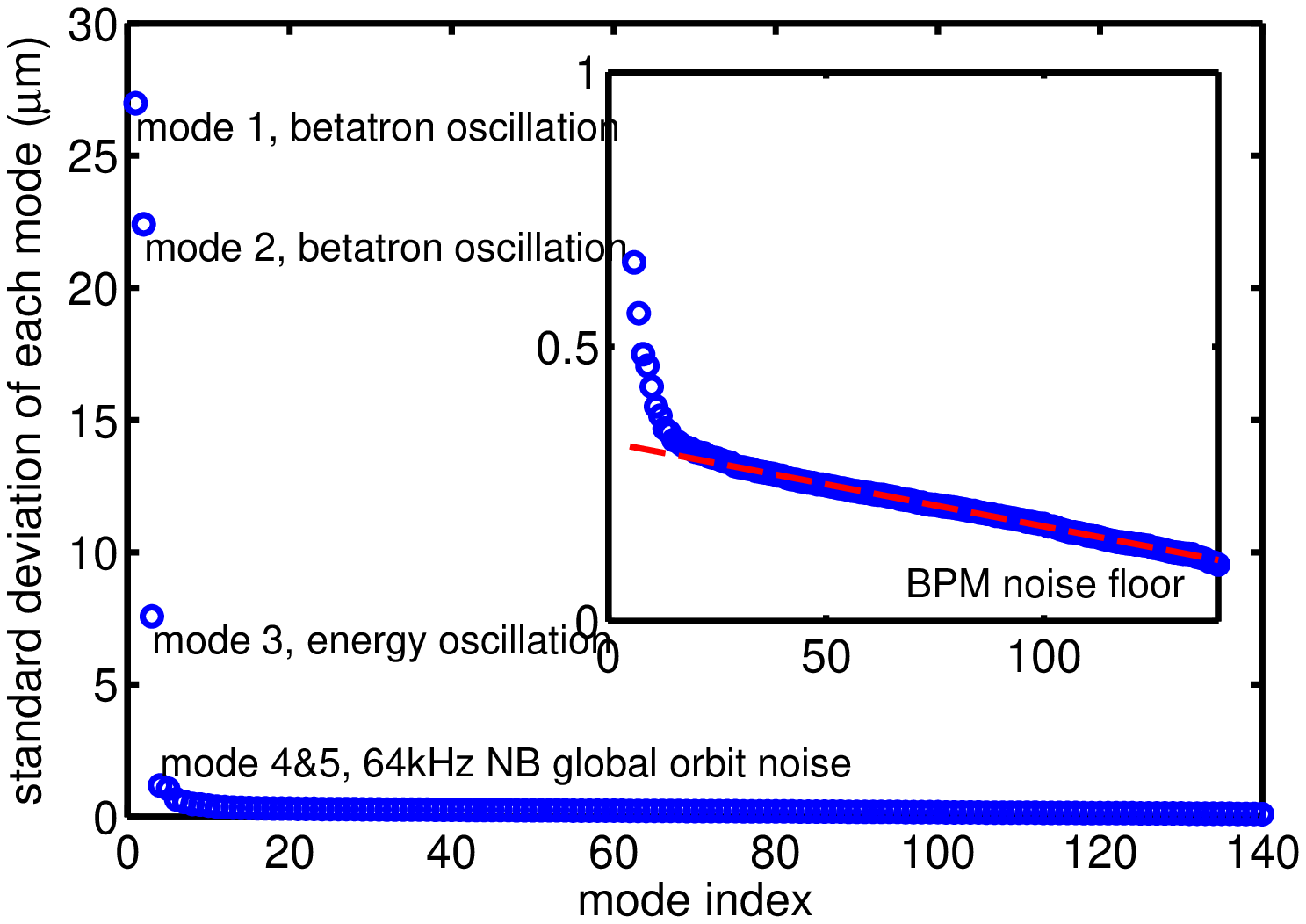}
\figcaption{\label{fig:typical-singular-values}Singular value plot of a typical SVD result. The noise floor reflects the average performance of global BPM system.}
\end{minipage}\end{center}

The first two modes turn out to be the well-known betatron oscillation and the third mode the energy oscillation after analyzing the spatial and the temporal matrices.\cite{Wang:2003:PAC:1,Wang:2003:PAC:2,Wang-Sajaev-Yao:2003:PRSTAB} Figures~\ref{fig:spectra-of-modes-12345}, \ref{fig:beta-function} and~\ref{fig:eta-function} show that the behaviors---temporal waveforms, frequencies and the corresponding amplitudes, conformity relations between the spatial vectors and the corresponding Twiss parameters, and even the singular point at BPM number 68---of the first three modes do fit characteristics of the $\beta$\nobreakdash-function and the $\eta$\nobreakdash-function.

\begin{center}\begin{minipage}{\hsize}
\includegraphics[width=\hsize]{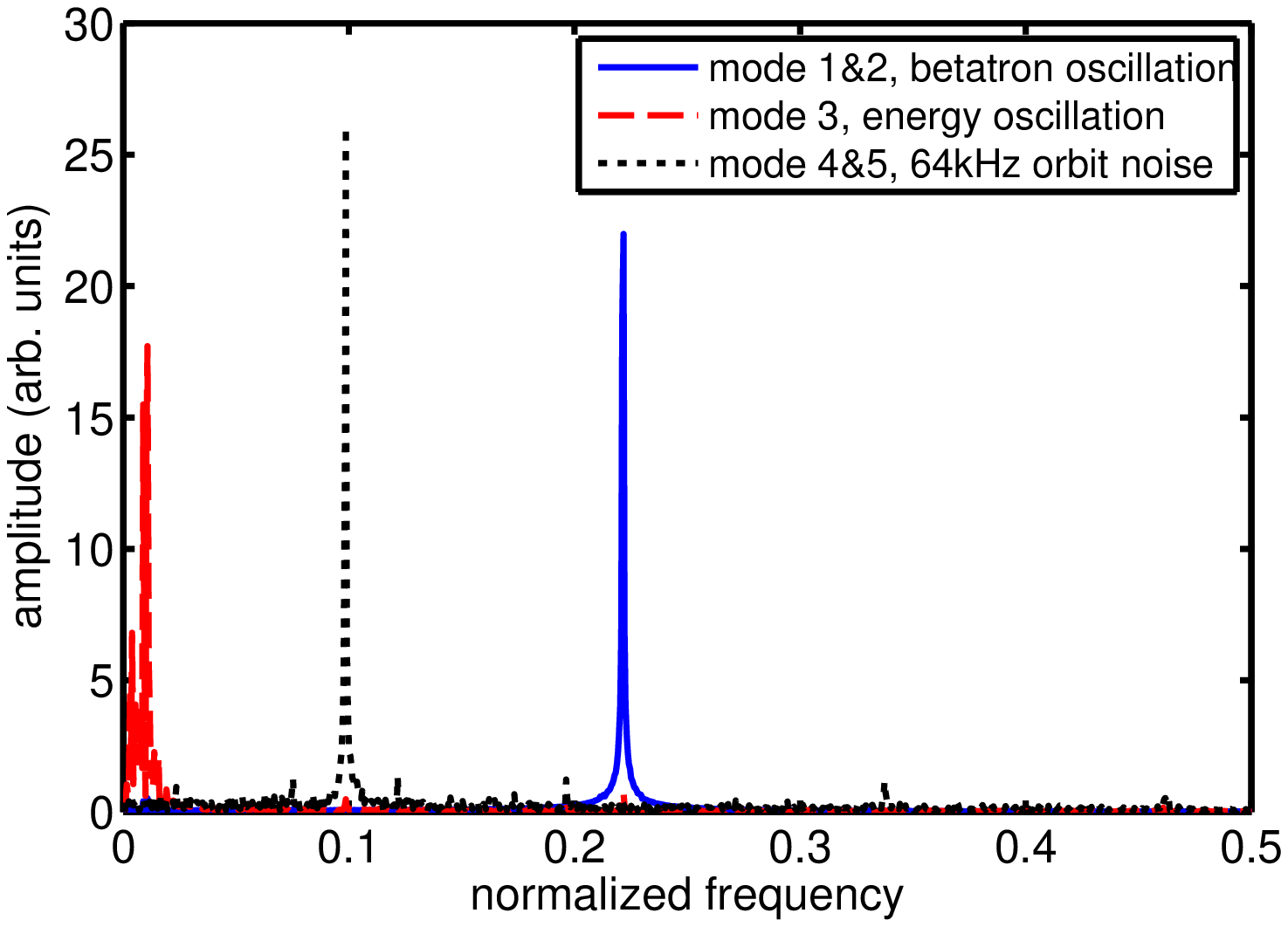}
\figcaption{\label{fig:spectra-of-modes-12345}Spectra of the first five modes. The first and second modes share the same spectrum with only the phase difference of $\pi/2$, and therefore are believed the bases of a single physical mode, as well as the fourth and fifth modes.}
\end{minipage}\end{center}

\begin{center}\begin{minipage}{\hsize}
\includegraphics[width=\hsize]{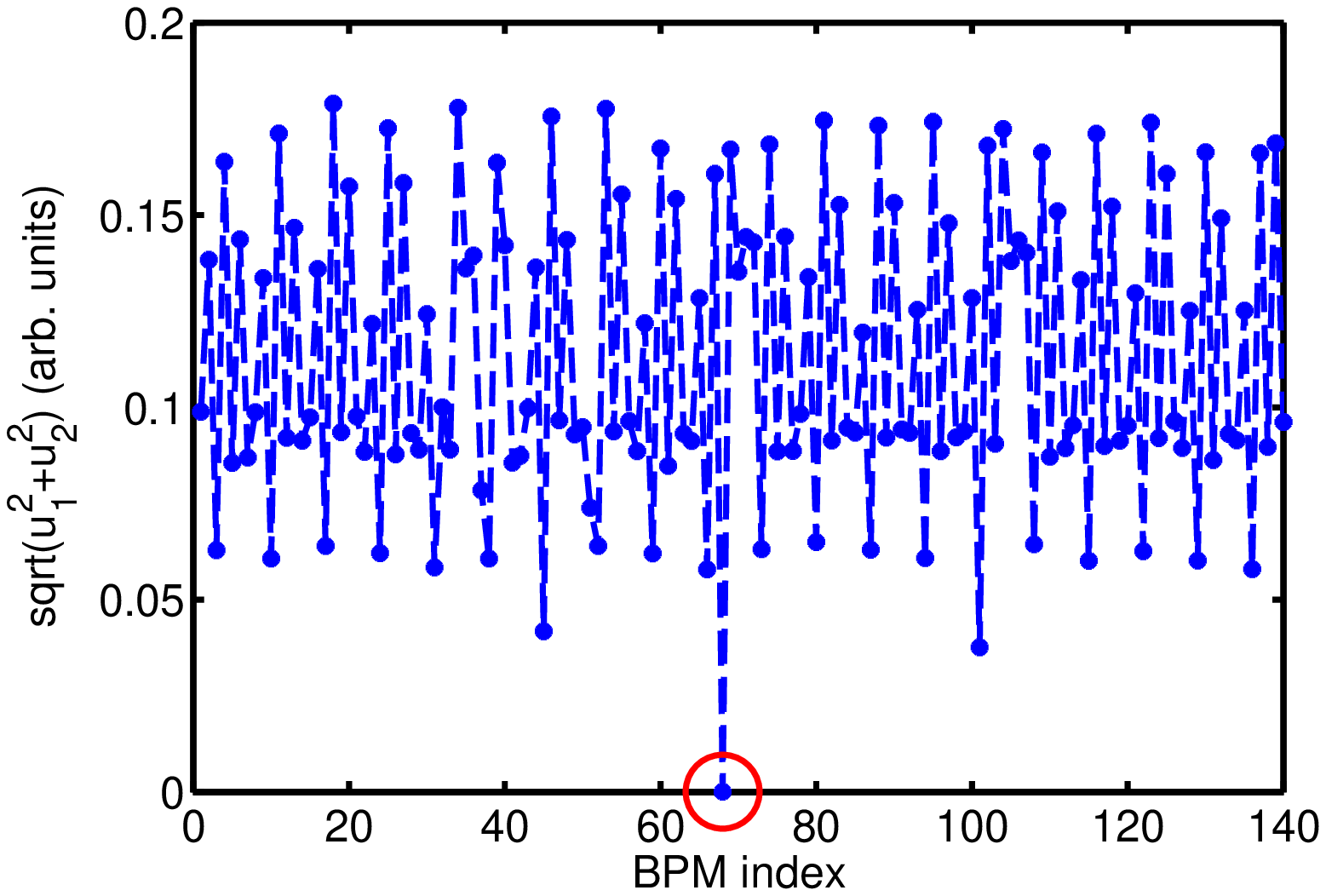}
\figcaption{\label{fig:beta-function}Spatial vector of mode~(\#1,\#2), which can be identified to be the betatron oscillation. The root of the quadratic sum of the first two spatial vectors is proportional to the $\beta$-function of the ring. The red circle marks the bad BPM.}
\end{minipage}\end{center}

\begin{center}\begin{minipage}{\hsize}
\includegraphics[width=\hsize]{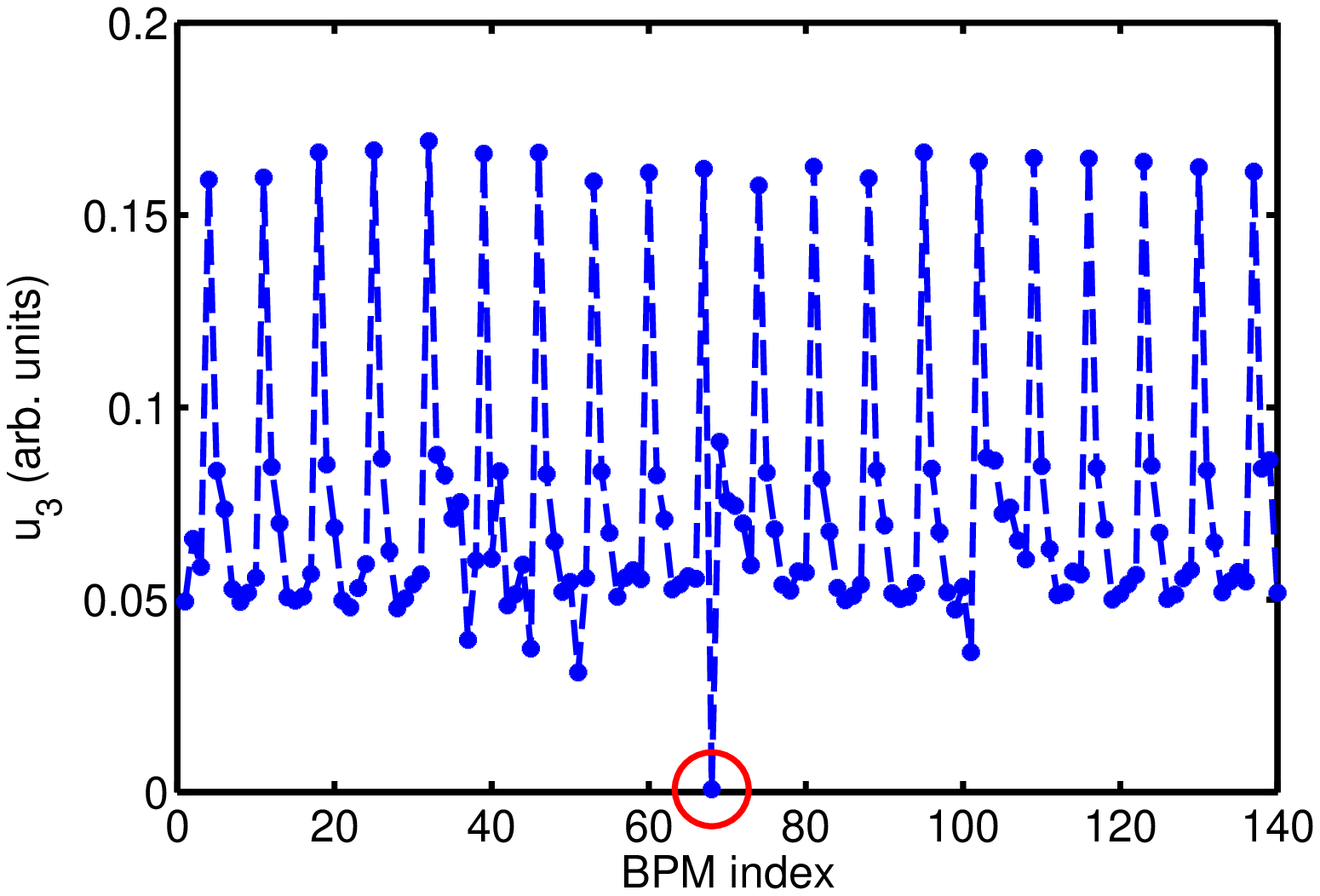}
\figcaption{\label{fig:eta-function}Spatial vector of mode~\#3, which can be identified to be the energy oscillation. The spatial vector is proportional to the dispersion function of the ring. The red circle marks the bad BPM.}
\end{minipage}\end{center}

After eliminating the first five modes which can be explained as the beam dynamics, the terms that remain would be considered measurement noise. Equation~\eqref{eq:SVD-Einstein} thus can be rewritten as:
\begin{equation}\label{eq:SVD-Einstein-components}
b_n^t = u_\beta^t s_\beta^\beta v_n^\beta + u_\eta^t s_\eta^\eta v_n^\eta + u_{GN}^t s_{GN}^{GN} v_n^{GN} + u_N^t s_N^N v_n^N ,
\end{equation}
where $\beta$ denotes the betatron oscillation, $\eta$ the energy oscillation, $GN$ the global narrow-band beam orbit noise and $N$ the noise of the BPM system. The matrix form of equation~\eqref{eq:SVD-Einstein-components} is
\begin{equation}\label{eq:SVD-components}
B = U(S_{1,2}+S_3+S_{4,5}+S')V^\dagger ,
\end{equation}
where $S_{n}$ only contains the $n$th diagonal element of $S$ and $S'=S-S_{1,2}-S_3-S_{4,5}$. Since the matrix
\begin{equation}\label{eq:noise-matrix}
B'=US'V^\dagger
\end{equation}
is now beam-independent, it could be used the estimate the BPM performance.

\section{Applications}

\subsection{Performance evaluation of global BPM system}

The resolution of the BPM system could be determined by finding the BPM noise floor of the diagonal elements of the singular matrix $S$ as shown in figure~\ref{fig:typical-singular-values}. Four sets of data to be analyzed are obtained by using different BPM electronics configurations due to the upgrading procedures in SSRF. Although the BPM system gave both horizontal data and vertical ones, only the horizontal ones are discussed here to avoid repetitions. The results in figure~\ref{fig:singular-values-stage} show that the floor had been split into two levels. The split implied that there were two different kinds of equipments with different resolutions. The numbers in the lower group coincided with the ones in table~\ref{tab:upgrade-details}. There are also intermediate states due to the entanglement of the noise modes. The horizontal resolutions of the global BPM systems with Libera Electron and Libera Brilliance are \SI{3.5}{\micro\metre} and \SI{1.1}{\micro\metre} respectively.

\begin{center}\begin{minipage}{\hsize}
\includegraphics[width=\hsize]{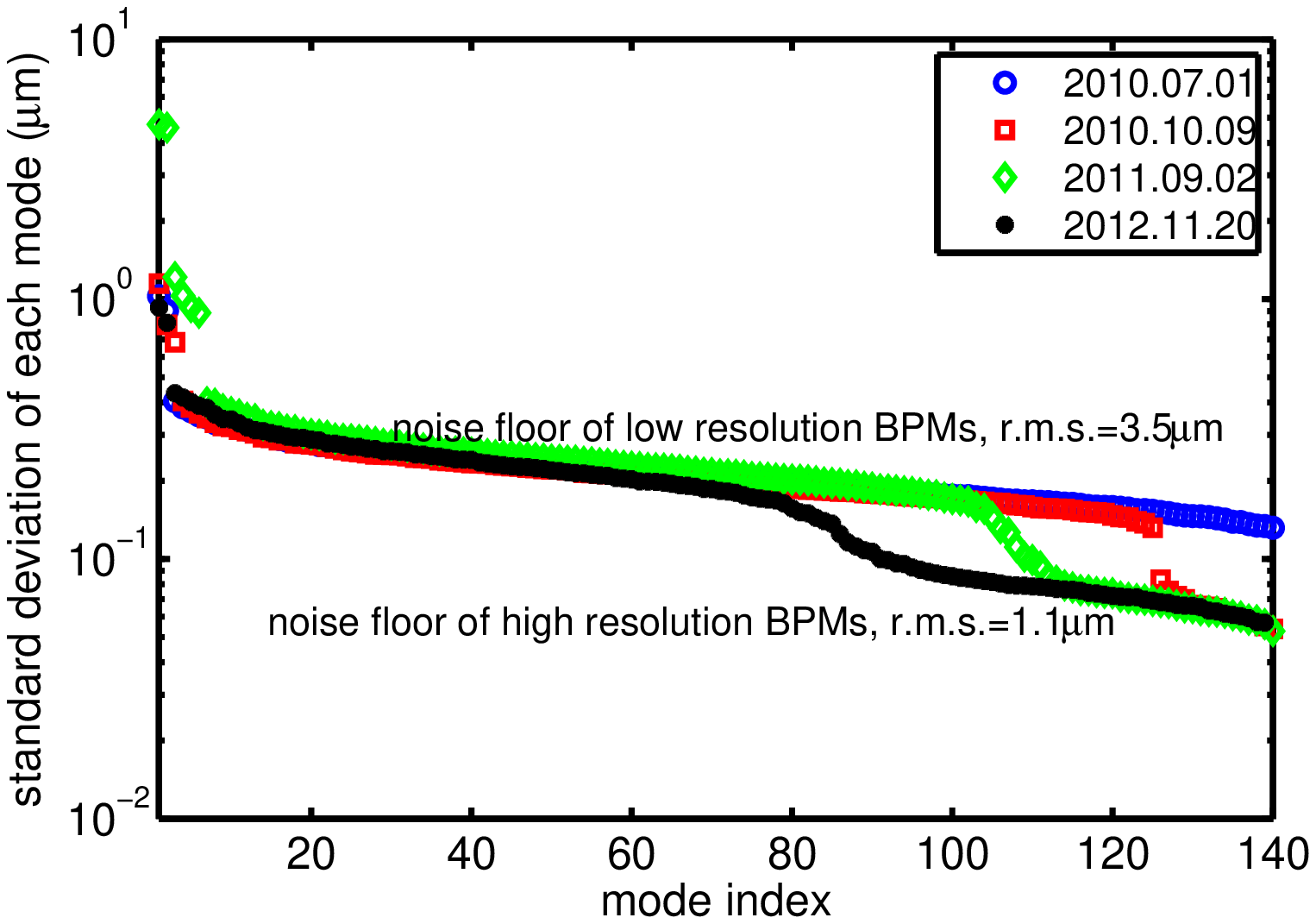}
\figcaption{\label{fig:singular-values-stage}Average performance evaluation for different upgrading phases using singular value plots. The upgraded BPMs are separated from the old ones.}
\end{minipage}\end{center}

\subsection{Performance evaluation of individual BPM}

Each column of the noise matrix \eqref{eq:noise-matrix} contains the random noise recorded by the corresponding BPM, so that the deviation of the column waveform indicates the resolution of the BPM. After calculating the standard deviations of the BPMs in different upgrading phases, the changes of the performance of individual BPM were studied. The BPMs upgraded in table~\ref{tab:upgrade-details} are marked black in figure~\ref{fig:upgrade-stage} and they do have qualitatively better performances than the old electronics as they were claimed.

Using the method routinely, one could easily calculate the vertical resolutions. Figure~\ref{fig:SVD-resolution} shows the results of the Libera Electron~(in drift sections or arc sections) and Libera Brilliance. It can be seen that both the resolution and the stability of the Libera Brilliance electronics are significantly improved.

\begin{center}\begin{minipage}{\hsize}
\includegraphics[width=\hsize]{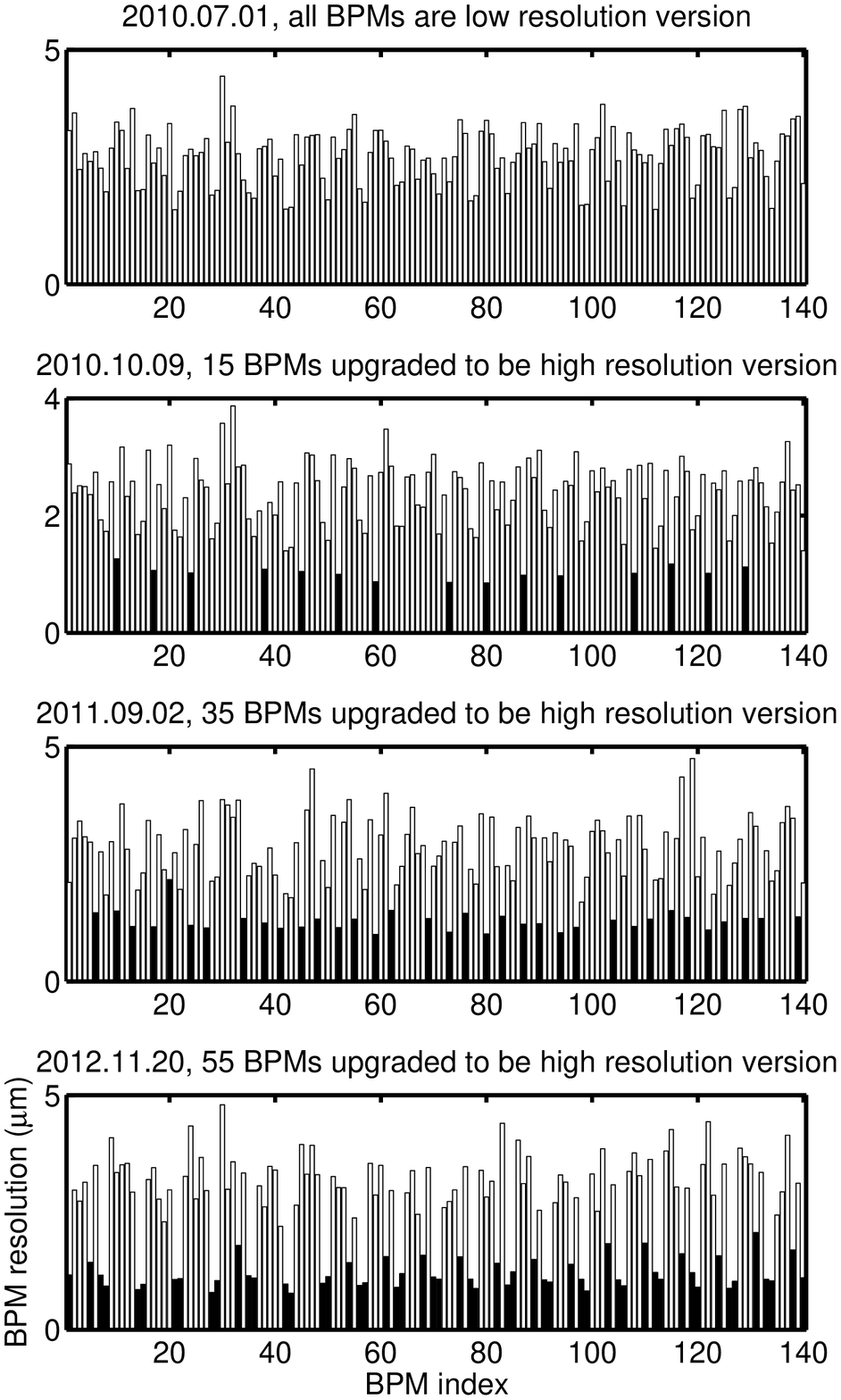}
\figcaption{\label{fig:upgrade-stage}Spatial resolutions of individual BPM at different upgrading phases. Dark bars indicate the upgraded BPMs.}
\end{minipage}\end{center}

\begin{center}\begin{minipage}{\hsize}
\includegraphics[width=\hsize]{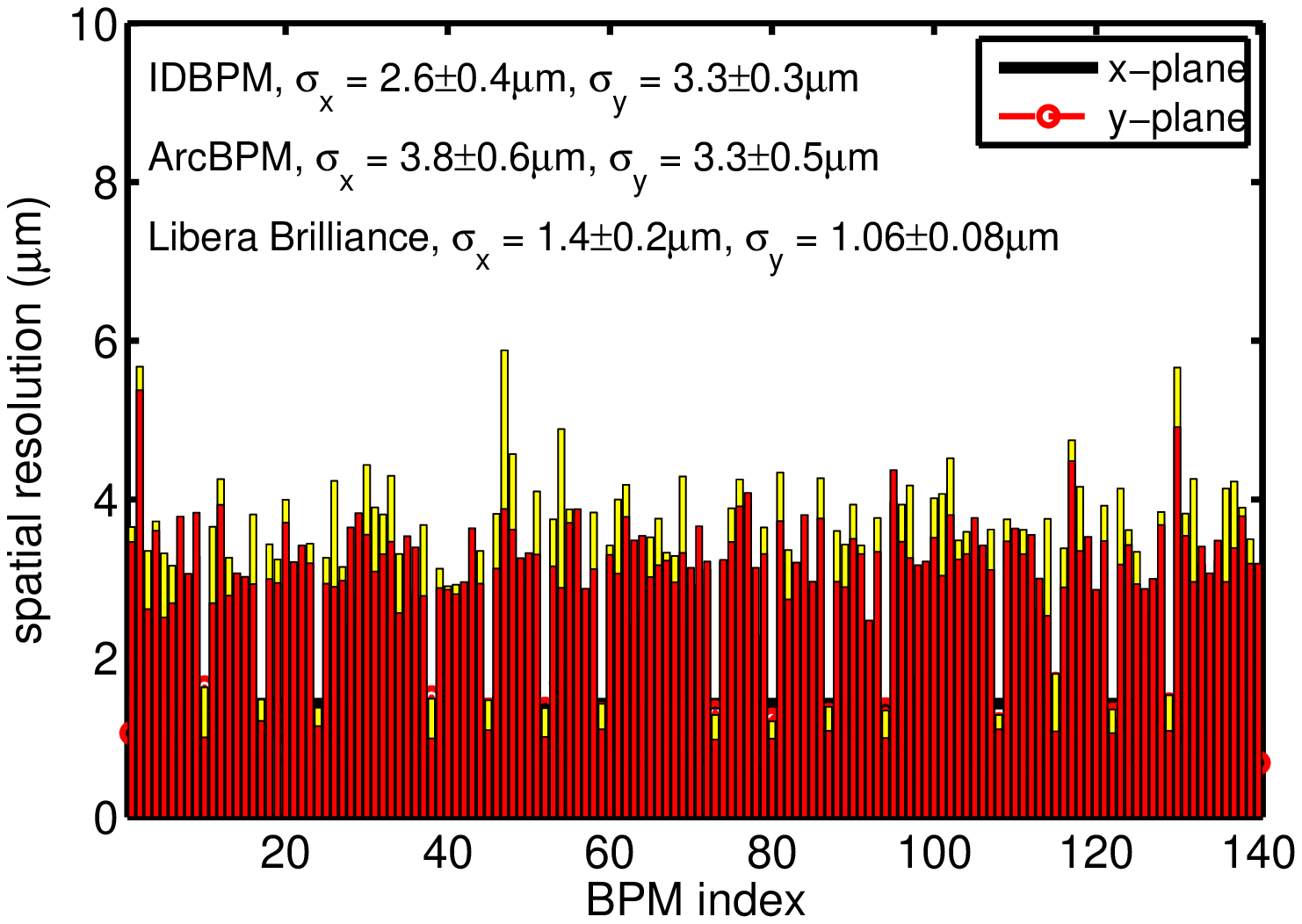}
\figcaption{\label{fig:SVD-resolution}Horizontal and vertical BPM resolution results.}
\end{minipage}\end{center}

\subsection{Beam status dependency}

The resolution of the BPM is actually related to the beam current due to signal-noise ratio. The relation between the beam current and the corresponding BPM global resolution can be calculated by using the method discussed in the previous section. Figure~\ref{fig:current-resolution} shows that the resolution and stability of the system kept improving while the beam current was increasing. A comparison in figure~\ref{fig:current-resolution-compare} had been made to show that without the elimination of the beam dynamics, the standard deviation of the raw data not only underestimated the BPM performance, but also saw the beam instability~(there was some kind of current related instability which had the threshold of \SI{200}{\milli\ampere}). These beam dynamics and instabilities were the actual signals and should not be regarded as part of the random noise.

\begin{center}\begin{minipage}{\hsize}
\includegraphics[width=\hsize]{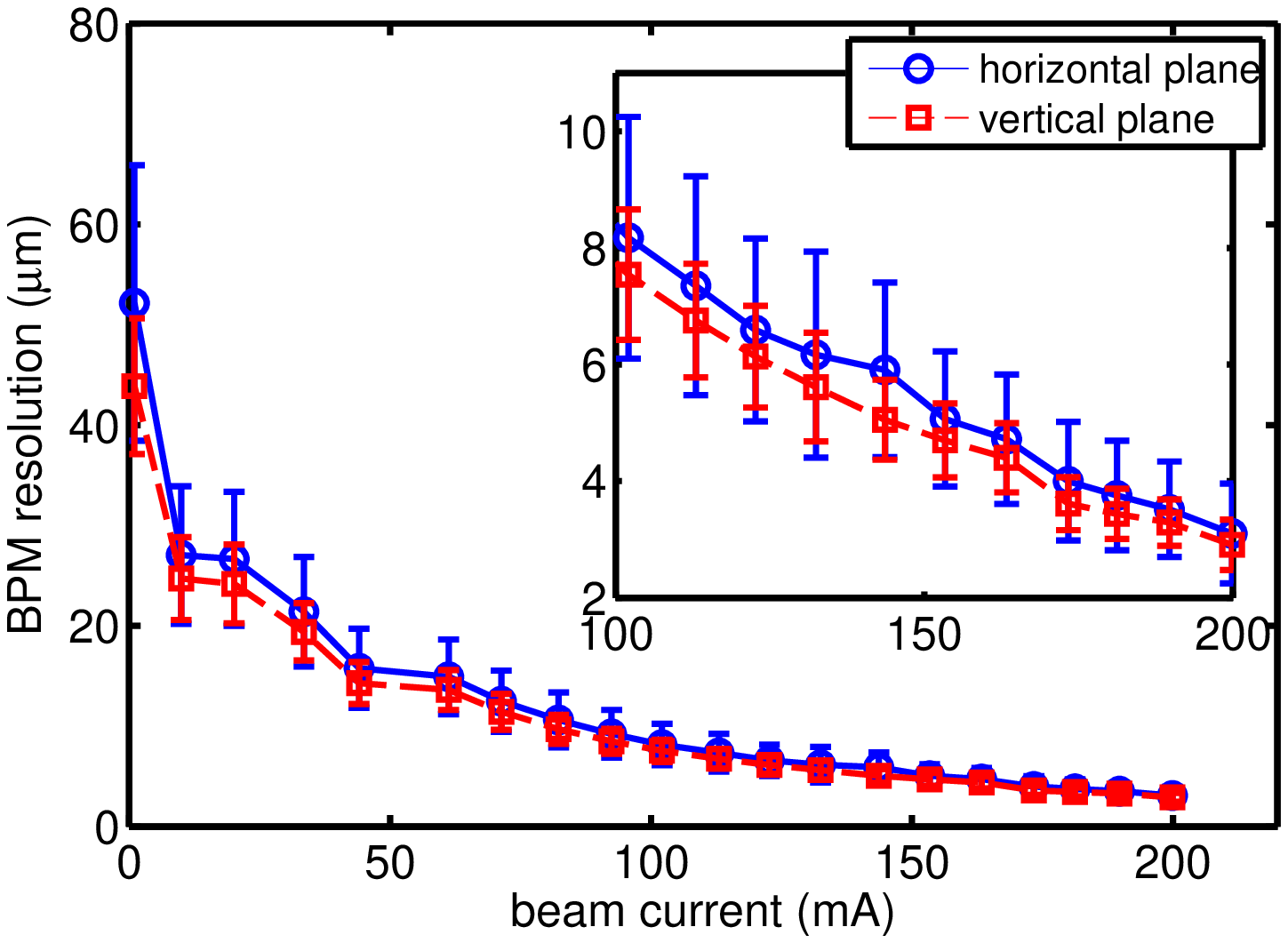}
\figcaption{\label{fig:current-resolution}Resolutions under different currents.}
\end{minipage}\end{center}

\begin{center}\begin{minipage}{\hsize}
\includegraphics[width=\hsize]{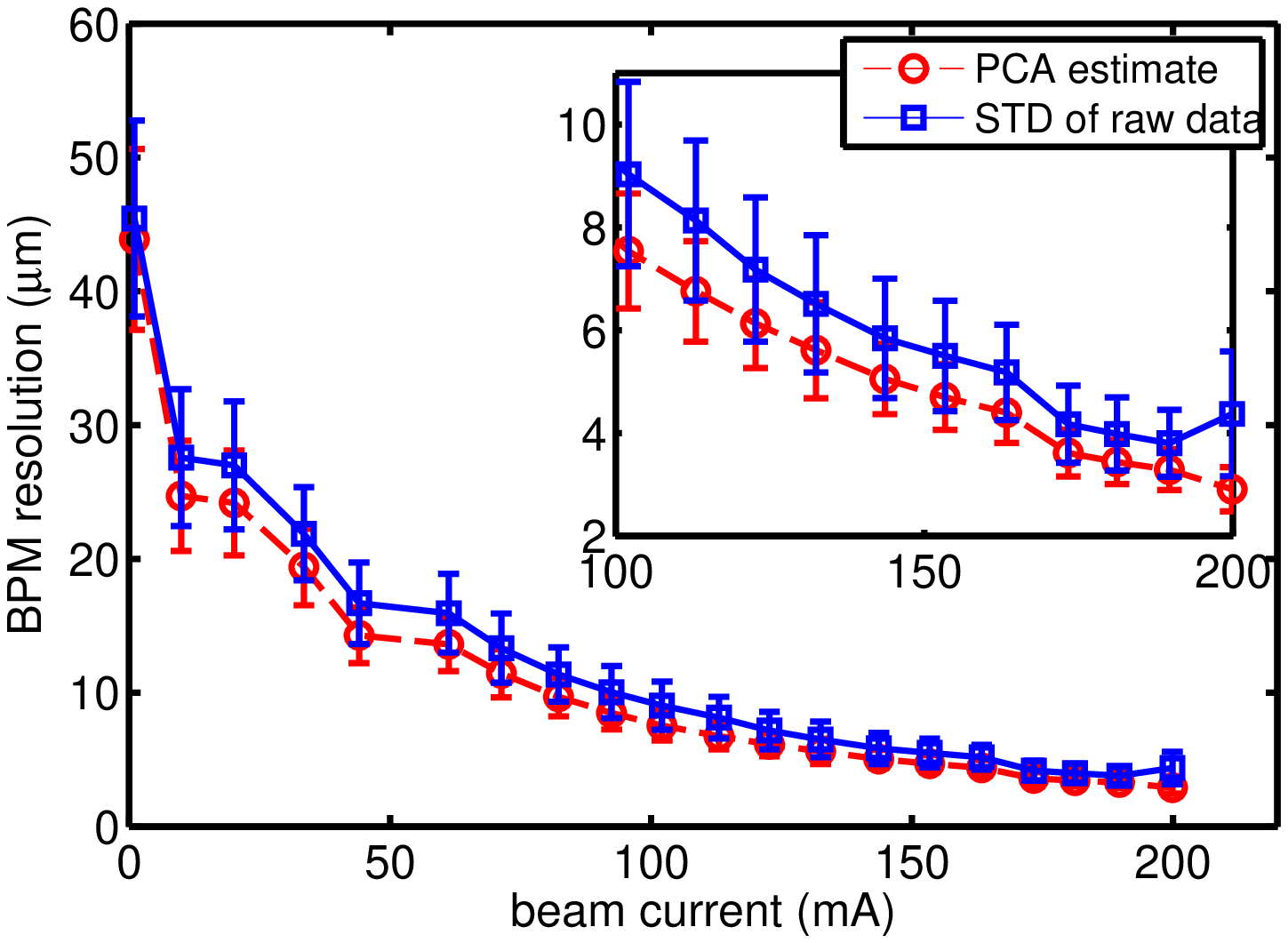}
\figcaption{\label{fig:current-resolution-compare}Comparison between the SVD and the standard deviation results of the BPM performance.}
\end{minipage}\end{center}

Another experiment was made to check whether the operation status could influence the BPM resolution. The transverse feedback and the injection kicker were switched to ``on'' or ``off,'' and the physical modes were affected, e.g., the betatron oscillation was suppressed or invoked in different situations, as could be seen in figure~\ref{fig:singular-value-operations}. Nevertheless, the noise floor did not seem to be affected and the pattern of the split kept unchanged. Therefore, the results of the BPM resolution would be the same in spite of the machine status. Figure~\ref{fig:relative-sigma} shows that $\pm10\%$ accuracy would be guaranteed if the injection kicker was turned on and $\pm20\%$ accuracy would be assured if the transverse feedback system was also turned off, for most BPMs.

\begin{center}\begin{minipage}{\hsize}
\includegraphics[width=\hsize]{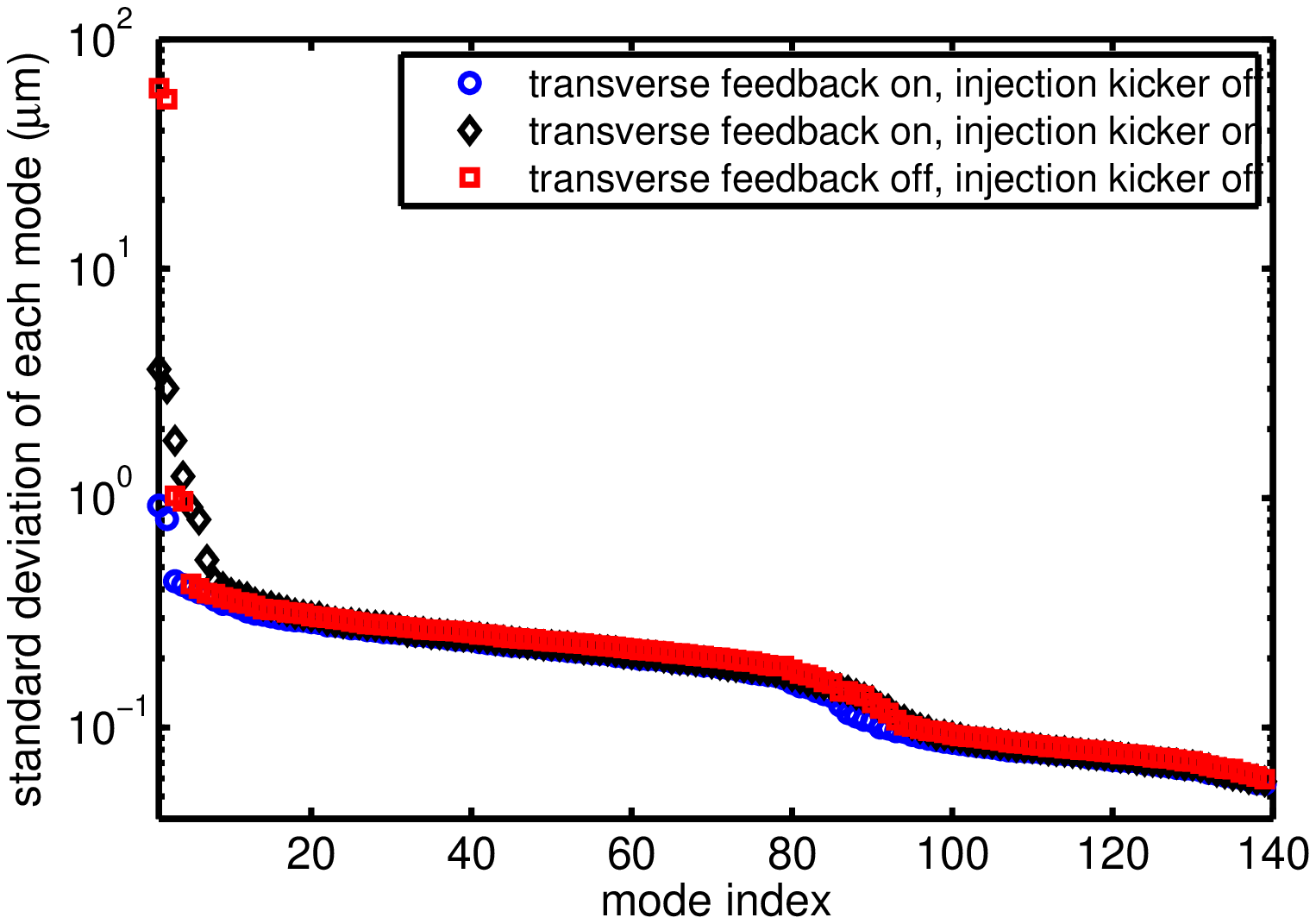}
\figcaption{\label{fig:singular-value-operations}Singular value plots of different data sets.}
\end{minipage}\end{center}

\begin{center}\begin{minipage}{\hsize}
\includegraphics[width=\hsize]{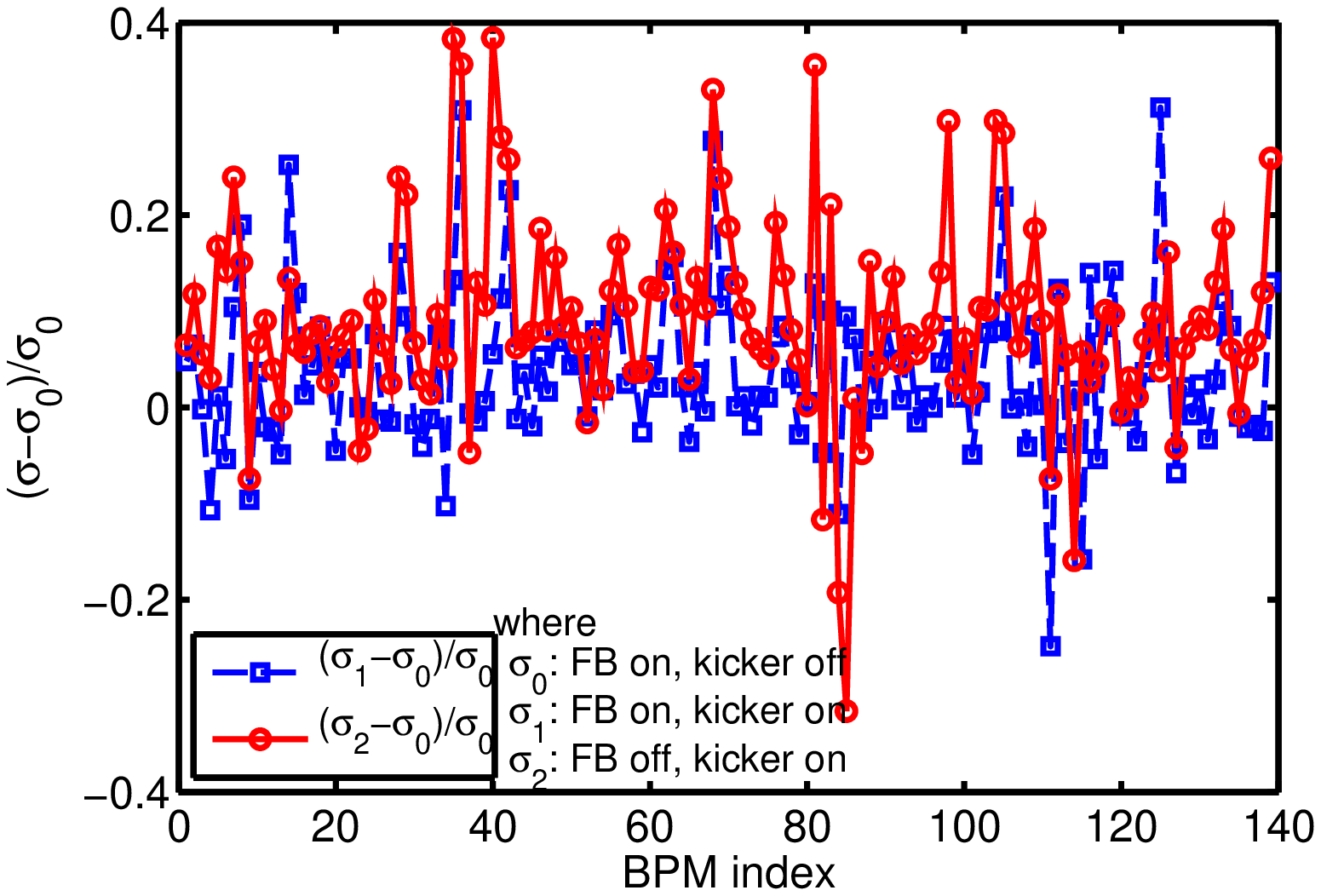}
\figcaption{\label{fig:relative-sigma}Relative resolution calculation deviations of each BPM.}
\end{minipage}\end{center}

\subsection{BPM configuration dependency}

The resolutions calculated in the previous sections were believed to be the electronics noise. The geometrical layout coefficients, denoted by $k_{x,y}$, were used to verify the assumption. These coefficients were determined by the coordinates of the probes and would be used to process the output signals. The resolution of the electronics noise is proportional to $k$ but the coefficient is unknown and varies from BPM to BPM.

Fortunately, the horizontal and vertical $k$ settings were not identical for all BPMs and the ratio $\sigma_x/\sigma_y$ could be used to cancel the irrelevant coefficient to see if it matched $k_x/k_y$~(as shown in figure~\ref{fig:k-ratio}). The $(k_x,k_y)$ values of the ID BPMs and the arc BPMs are $(\SI{12.1}{\milli\metre},\SI{11.9}{\milli\metre})$ and $(\SI{19.1}{\milli\metre},\SI{13.4}{\milli\metre})$, respectively. There were still beam related components and alignment errors, so that some of the BPMs did not get a perfect match, but the pattern could be confirmed within reasonable amounts of errors. On the other hand, the results of the standard deviations of the raw data did not show the pattern and the ratios of $\sigma_x$ and $\sigma_y$ were dominated by the horizontal oscillation, despite the transverse feedback suppression of the betatron motion, so that they were relatively larger than the theoretical values.

\begin{center}\begin{minipage}{\hsize}
\includegraphics[width=\hsize]{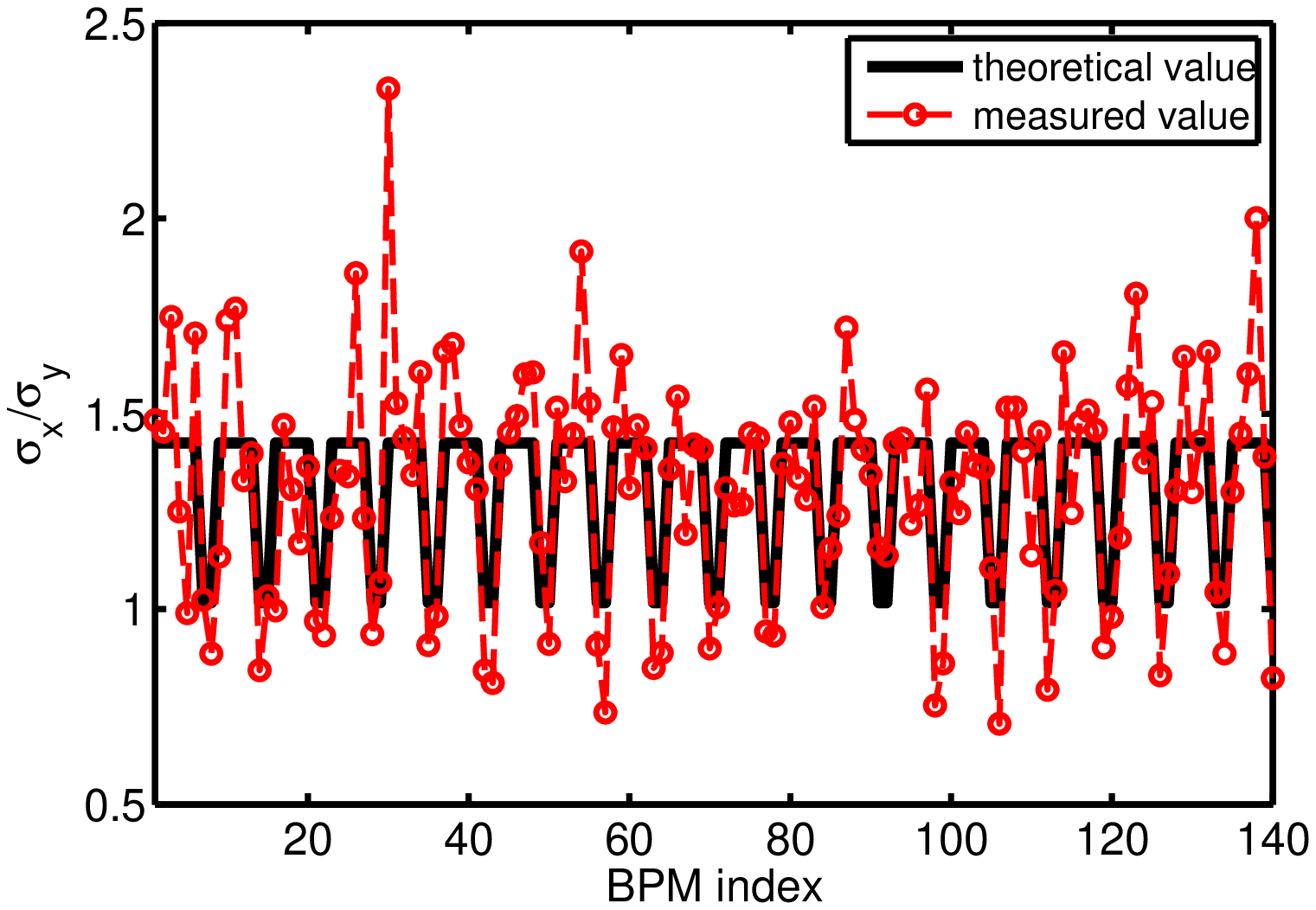}
\figcaption{\label{fig:k-ratio}Comparison between $\sigma_x/\sigma_y$ and its theoretical value $k_x/k_y$.}
\end{minipage}\end{center}

\section{Conclusion}

PCA is good at separating linear-independent signals which is exactly what's needed in the resolution estimate. The main physical modes had been successfully extracted and identified from the raw data by using the SVD to decompose various of signals, and the rest modes were considered the instability of the measurement system. The beam based resolution measurement of the BPM system then was done by using the noise matrix.

Experiments had shown that PCA not only could be used to discover the beam dynamics signals, but also was capable of distinguishing one type of electronics noise from another by the singular values. Further more, the characteristics of the horizontal and vertical resolutions met the geometrical coefficients of the BPM probe. Those experiments had proved that the validity of this method.

\end{multicols}

\end{document}